# FINDING CORE MEMBERS OF COOPERATIVE GAMES USING AGENT-BASED MODELING

Daniele Vernon-Bido, and Andrew J. Collins

Old Dominion Unversity

Norfolk, Virginia, USA

ajcollin@odu.edu

## ABSTRACT

Agent-based modeling (ABM) is a powerful paradigm to gain insight into social phenomena. One area that ABM has rarely been applied is coalition formation. Traditionally, coalition formation is modelled using cooperative game theory. In this paper, a heuristic algorithm is developed that can be embedded into an ABM to allow the agents to find coalition. The resultant coalition structures are comparable to those found by cooperative game theory solution approaches, specifically, the core. A heuristic approach is required due to the computational complexity of finding a cooperative game theory solution which limits its application to about only a score of agents. The ABM paradigm provides a platform in which simple rules and interactions between agents can produce a macro-level effect without the large computational requirements. As such, it can be an effective means for approximating cooperative game solutions for large numbers of agents. Our heuristic algorithm combines agent-based modeling and cooperative game theory to help find agent partitions that are members of a games' core solution. The accuracy of our heuristic algorithm can be determined by comparing its outcomes to the actual core solutions. This comparison achieved by developing an experiment that uses a specific example of a cooperative game called the glove game. The glove game is a type of exchange economy game. Finding the traditional cooperative game theory solutions is computationally intensive for large numbers of players because each possible partition must be compared to each possible coalition to determine the core set; hence our experiment only considers games of up to nine players. The results indicate that our heuristic approach achieves a core solution over 90% of the time for the games considered in our experiment.

# INTRODUCTION

Agent-based modeling (ABM) is a frequently used paradigm for social simulation (Axtell, 2000; Epstein & Axtell, 1997; Gilbert, 2008); however, there is little evidence of its use in coalition formations especially strategic coalition formations. There are few models that explore coalition formation (Collins & Frydenlund, 2018; Sie, Sloep, & Bitter-Rijpkema, 2014) and even fewer that validate their results against an expected outcome (Abdollahian, Yang, & Nelson, 2013). Cooperative game theory is often used to study strategic coalition formation, but solving games involving a significant number of agents is computationally intractable (Chalkiadakis, Elkind, & Wooldridge, 2011a). The ABM paradigm provides a platform in which simple rules and interactions between agents can produce a macro-level effect without the large computational requirements. As such, it has the potential to be an effective means for approximating cooperative game solutions for large numbers of agents. In this paper, we intend to show one possible approach to approximating strategic coalition formations in an agent-based model. The model is validated through a comparison of its outputs to those from cooperative game theory.

Cooperative game theory provides an analytic framework for the study of strategic coalition formation (Chalkiadakis, Elkind, & Wooldridge, 2011b). Given a situation where a decision-maker must evaluate which coalitions they wish to join, cooperative game theory aids in this evaluation. Games that are expressed in terms of the decision-makers' (or players') preference for different coalitions are called hedonic games. A common example of an hedonic game is the roommate problem (Aziz, 2013; Gale & Shapley, 1962). Here is a simple example of a classic roommate problem: Campus housing is expensive, so students might seek a roommate rather than live alone; a group of three students (A, B, and C) has found a two-bedroom apartment. Student A prefers to room with student B over student C. Student B prefers student C over student A. Student C prefers student A over student B. No student wants to share the apartment with two roommates as one would have to sleep in the living room. Each student demonstrates a preference for a different coalition leading to a cycle of dissatisfaction. In game theory, the term "players" is used to refer to agents; as such, we use these names interchangeably throughout the paper.

The outcome of a cooperative game, including hedonic games, is a coalition structure and a payoff vector. A coalition structure is a partition of the players in the game, which is equivalent to a collection of disjoint coalitions that covers all players; the payoff vector is the utility received by the players. In a hedonic game, the utility, for a given player, is a function of the identity of the members in the coalition (Iehlé, 2007). There are several ways the solution set of cooperative games can be defined. We will use core stability as our solution concept (Bogomolnaia & Jackson, 2002). "A coalition structure is called stable if there is no group of individuals who can all be better off by forming a new deviating coalition. The core of a hedonic game is the collection of all stable coalition structures" (Sung & Dimitrov, 2007). Determining the core of a hedonic game is NP-complete (Ballester, 2004); for all but the smallest number of agents, it requires an unreasonable amount of computational time to solve. Additionally, if the core exists, it may contain any number of coalition structures. Therefore, a complementary problem for determining the core is to test whether a coalition structure is a member of the core. Testing core membership requires selection of a candidate coalition structure and determining whether the coalition structure fits the criteria to be a part of the core. Core membership testing is co-NP complete (Faigle, Kern, Fekete, & Hochstättler, 1997; Sung & Dimitrov, 2007). That is, it is equally as difficult to determine core membership as it is to find the core.

In this paper, we present an Agent-based model (ABM) for finding a core stable coalition structures; that is, coalition structures that are member of the core. Our model's algorithm considers a number of different ways to test a coalition structure by considering possible other coalitions that could form. It tests mergers of coalitions, adding an agent to a coalition, dividing a coalition, removing an agent from a coalition, agents breaking away to form a new coalition, and the individual rationality of being in the coalition. Agents examine each potential new coalition to determine if their utility is improved; if the utility is improved for all agents, the new coalition is accepted. Based on the cooperative game theory underpinnings, these procedures allow for the efficient estimation of a stable coalition structure and bring a strategic component to the ABM.

ABM is a modeling paradigm particularly well suited to handle the interactive nature of coalition formation and to, stochastically, surmise a feasible coalition structure. It is a computation method in which researchers can build and analyze models comprised of agents (players) that interact in an environment (Gilbert, 2008). While cooperative game theory mathematically evaluates the formation of every possible coalition, our ABM approach stochastically explores coalitions generated from agent interactions. The benefit of exploring stochastically-generated coalitions is the reduction of computational complexity. Generating every possible coalition structure for cooperative game theory analysis using dynamic programming has a computational complexity of *O($3^n$)* (Rothkopf, Pekeč, & Harstad, 1998; Yeh, 1986) with the actual finding of the core set being NP-Complete for hedonic games (Ballester, 2004) while our ABM approach magnitude is linear per time-step  so the computational complexity is determined by the number of time-steps allowed. We implement six basic coalition formation behaviors with each selecting one player as the focal point thus reducing the computational complexity by not tying the algorithm run time to an exhaustive search. In this paper, we validate the success of our algorithm by solving the core of cooperative games using a 'brute-force' method, then comparing the results to those generated by the ABM algorithm for the same games. Our ABM algorithm produces core members, over 95% of the time.

Both cooperative game theory and agent-based modeling involve modeling multiple agents that are interacting; as such, it might be expected that there would be an overlap of the approaches, i.e., hybrid modeling (Mustafee et al., 2015). However, there is very little in the literature showing a combination of these modeling methods. Abdollahian et al. (2013) incorporated cooperative game theory concepts into an agent-based model of location siting for sustainable energy; they employ the Shapley value (Shapley, 1953) as their solution concept whereas we use the core (Gillies, 1959).  Ohdaira and Terano (2009) discuss cooperation in an agent-based simulation of the classic Prisoner's dilemma but they are focused on non-cooperative  game theory. Sie et al. (2014) discuss coalition formation in the context of social networks; they employ the Shapley value. Of the extant literature, the work presented in this paper relates to Collins and Frydenlund (2018), who created a hybrid model of cooperative game theory and agent-based modeling; their work will be discussed later in this paper.

Janovsky and DeLoach (2016) claim to have found a heuristic for solving cooperative games using agent-based modeling. Their approach focuses on a particular form of a characteristic function used in Transferrable Utility (TU) games. Our approach focuses on hedonic games which are a type of non-transferrable Utility (NTU) game.

The remainder of the paper is formatted in the following manner. Section two provides a background for those not familiar with hedonic games. Section three provides our ABM algorithm for determining a coalition structure. Section four describes a "brute-force" method of solving cooperative games. This method is used to determine the core of a cooperative game given ample time and resources. This is followed by a use case and the results. Finally, we conclude the paper with a summary of our work.

## COOPERATIVE GAME THEORY AND HEDONIC GAMES

Game theory is "the study of mathematical models of conflict and cooperation between intelligent, rational decision-makers" (Myerson, 2013, p. 1). Game theory can be split into non-cooperative and cooperative games. Non-cooperative games usually involve only two rational players and they are well studied (Eatwell, Milgate, & Newman, 1987; Von Neumann & Morgenstern, 1947). Prisoner's Dilemma, named by Albert Tucker in 1950, is possibly the best known non-cooperative game (Flood, 1952). Games involving more than two players are usually studied using n-person game theory which is also known as cooperative games (Chakravarty, Mitra, & Sarkar, 2015). In non-cooperative games, both parties know the potential payoff based on the choices made but they do not have the ability to create a binding agreement before a course of action is selected. Cooperative games, on the other hand, are games in which players can enter into contracts or binding agreements that allow them to form strategic coalitions (Peleg & Sudhölter, 2007).

Cooperative games are generally divided into two types: games with transferrable utilities (TU) (also known as characteristic function games) and games with non-transferrable utilities (NTU) (Thomas, 2003). Characteristic function games allow a coalition to subdivide the payoff, obtained by the coalition, in any possible way (Peleg & Sudhölter, 2007); this introduces a two-stage problem for the game; namely, you need to determine which coalitions will form first and how those coalitions will subdivide the payoff amongst the coalition's members. In contrast, NTU games have fixed distributions of the payoff dependent upon the coalition, for example, a member of coalitions payoff might be their preference for being in one coalition and, as such, this payoff cannot be transferred to another player (e.g., I cannot give my joy of being admitted to an exclusive club to another member of that club). Hedonic games are a specific type of NTU game were players have a preference relation over all possible coalitions (Aziz & Savani, 2016). Hedonic coalitions, the term originally credited to Dreze and Greenberg (1980), describes how and why various groups, clubs, and communities form and persist (Iehlé, 2007). Technically, a coalition in a NTU game could have multiple choices of action to determine their payoff; a hedonic game is NTU game when there is only one course of action for each coalition (Chalkiadakis et al., 2011b). Since payoffs are fixed for a hedonic game, solving is only a one-stage problem, i.e., you just need to determine which coalitions will form.

The formal structure of a hedonic game is defined as $G = (N, \succsim_1, \ldots, \succsim_n)$ where $N = \{1, \ldots, n\}$ is the non-empty set of players in the game, and $\succsim_i$ is the preference relationship of possible coalitions containing $i \in N$. Chalkiadakis et al. (2011a) define the preference relation as a binary function with properties of completeness, reflexivity, and transitivity. Completeness denotes that for every pair of coalitions containing $i$, $c_j^i, c_k^i \subseteq N$, there is a relationship of $c_j^i \succsim_i c_k^i$ or $c_k^i \succsim_i c_j^i$. Reflexivity represents the fact that $c_j^i \succsim_i c_j^i$. The transitivity properties state that for $c_l^i \subseteq N, if\ c_j^i \succsim_i c_k^i\ and\ c_k^i \succsim_i c_l^i\ then\ c_j^i \succsim_i c_l^i$. The outcome of a hedonic game is a coalition structure. A coalition structure is a partition of all players of the game; that is, it is a set of coalitions which contains every player in exactly one coalition; the union of the coalitions in the coalition structure is *N* and the intersection of any pair is the null set. The number of possible coalition structures is combinatorial to the number of players *(n)*. For example, a game with 15 players results in over 1.3 billion possible coalition structures (Rahwan, Ramchurn, Jennings, & Giovannucci, 2009). Since a coalition structure is equivalent to partition, the number of coalitions structures is a Bell number (Bell, 1938).

Not all coalition structures are equally desirable. Game theorists have developed solution concepts for cooperative games that identify sets of coalition structures with distinct properties. Bogomolnaia and Jackson (2002) define solution concepts for hedonic games: core stability, individual stability, Nash stability, and contractually individual stability. We focus on core stability, a coalition structure that is core stable is called a core partition (Banerjee, Konishi, & Sönmez, 2001). The foundation for the core stability concepts lies in Gillies (1959) definition of the core. The core partition is the set of coalition structures which are not dominated by any other coalitions. A coalition $C \subseteq N$ dominates (or blocks) a coalition structure *CS* if, for all $i \in C, C \succ_i CS_i$, where $CS_i$ is the coalition in CS that contains player i (Chalkiadakis et al., 2011a). The core partition, therefore, represents the set of coalition structures in which there is no incentive for any subset of players to create a new coalition. The core partition "has the same requirement as the core with coalition structure studied in standard cooperative game" (Iehlé, 2007); hence we use the terms interchangeably within this paper. A game is core stable if it has a non-empty core.

There are a number of other possible solution concepts that can have been used, namely: individually stability, Nash stability, and contractual individual stability (Bogomolnaia & Jackson, 2002). A coalition structure is individually stable if there is not a coalition in which all members would be better off by having another player join the coalition, including that new player. A coalition structure is Nash stable if no player would want to join an alternate coalition, in the current coalition structure, or be a coalition unto themself. Contractual individual stability exists if there is no coalition change that a player can make that is preferable to the coalition the player left and the coalition the player joins (Bogomolnaia & Jackson, 2002; Chalkiadakis et al., 2011a).

The solution concepts provide an approach to evaluating coalition formation. Foundationally, it provides the analytic framework for determining which coalition structures are stable. However, a challenge remains. There are a finite number of coalitions, $2^N - 1$, but the number of coalition structures and the number of comparisons required is computationally infeasible for large number of agents. For example, a game of 15 players has 33,000 possible coalitions but 1.3 billion potential coalition structures. To determine the core of such a game requires 43 trillion comparisons to be made. This would take 3.5 hours on a single 3.50 GHz core computer, assuming the check can be completed in a single computational step (which is unlikely). If there were 20 agents, this would take 500 years. Our agent-based simulation runs in less than an hour. Determining the core of a hedonic game with arbitrary preferences, preferences that allow ties, is NP-complete (Ballester, 2004). The analytical solution of the core is computationally intensive and stochastically guessing whether a coalition structure is or is not a member of the core is akin to finding a needle in a haystack. Hence the desire to have heuristical methods like the one outlined in this paper.

## ABM HEURISTIC COALITION STRUCTURE SELECTION

Coalition formation represents the willingness of players to interact and/or cooperate with one another. Cooperative game theory provides a strategic approach to evaluating coalition formation, but it is computationally infeasible for large numbers of players. Agent-based modeling (ABM) provides a platform for evaluating interactions of many heterogeneous agents but is often found to be lacking the strategic component (Collins & Frydenlund, 2018). Combining the strategic nature of cooperative games with the interaction paradigm of ABM will create a more computationally efficient way in which to explore coalition formation. Game-theoretic results can be enhanced by ABM (De Marchi & Page, 2008).

Interactions, or opportunities to form coalitions, can occur between individual players, existing coalitions, players, and existing coalitions or within coalitions. Rules of coalition formation within the interaction platform are defined by the hedonic game solution concept: for a new coalition to form, each player in the coalition must prefer the new coalition to its existing coalition. In this way, each coalition change should result in a more stable coalition structure than the predecessor. This, theoretically, should produce a coalition structure that is a member of the core, if the core is non-empty, without an exhaustive search of the solution space. The heuristic presented here advances the core membership concept by Collins and Frydenlund (2018).

Our ABM algorithm is strategic in its determination of whether to change coalitions, i.e., agents will only join a newly formed coalition if it increases their utility; however, it is also heuristic as it stochastically selection of which coalitions to test. Our algorithm consists of six routines for choosing a coalition to test: (1) Join coalitions; (2) Exit coalition; (3) Create pair coalition; (4) Defect coalition; (5) Split coalition; and (6) Return to an individual coalition. Each of these routines is talked about in turn.

*Join coalition*

Two agents from different coalitions are chosen randomly. The payoffs of the merged coalitions are calculated. If the payoffs improve for all members of both coalitions, a new coalition is formed, which is the merged coalition.

*Exit coalition*

An agent from a coalition whose size is greater than one, i.e., not a singleton coalition, is randomly selected. The payoff of the coalition minus the agent is calculated. If all agents in the remaining coalition improve their payoff by removing the selected agent, the agent is removed from the coalition and forms a singleton coalition.

*Create a pair coalition*

Two agents are randomly selected. The payoff for the coalition containing just both agents is calculated. If the payoff of both agents is improved in this new coalition, both agents exit the current coalition in favor of the new one.

*Defect coalition*

A randomly chosen agent selects a coalition to which he does not belong. If joining this coalition improves his payoff and the payoff of all members of the coalition, the agent defects from his current coalition and joins the new coalition.

*Split coalition*

A coalition is randomly chosen, and a random subset of agents from the coalition are selected to form a separate coalition. If the members of the new coalition improve their payoff or the coalition that remained improve their payoff, the coalition splits into the two coalitions.

*Return to an individual coalition*

An agent is randomly chosen. If that agent would be better off on their own, i.e., they prefer the singleton coalition to their current coalition; then, they leave their current coalition to form the singleton coalition. This is known as the individual rationality concept (Thomas, 2003).

Players in the game are assumed to be self-interested and individually rational (Chalkiadakis et al., 2011a); that is, they are driven by achieving their preferred coalition and will not join a coalition that does not at least provide satisfaction equal to what they receive being on their own. The final routine ensures that players will only remain in a coalition that is preferable to being alone. In our simulation runs, all agents start on their own.

The six routines are iterative; they are repeated until the coalition structure becomes stable (i.e., no additional changes occurring to the coalitions for a certain number of time-steps). The repetition allows players to attempt to migrate into higher preferred coalitions. However, the algorithm arriving at a stable state with respect to coalition structure does not guarantee core membership. It should be noted that not every player and coalition is considered at each timestep, which is a departure from the algorithm from which ours is based because it considered every player at every timestep (Collins & Frydenlund, 2018).

The Collins' and Frydenlund's algorithm was never formally compared to the actual cooperative game theory core set, though a primary analysis shows that it is not proficient at achieving core member coalition structures (in some cases it only reach the core solution 4% of the time, which is vastly worse than the results presented for our new algorithm). It examines all agents at every timestep and only considers a subgroup split and supergroup group formation. This original algorithm approach was applied to several areas including economic minority games (Collins, 2017), generic neighborhood alliances (Collins & Frydenlund, 2016b), and refugee movement (Collins & Frydenlund, 2016a). Our algorithm is designed with the intent of creating a coalition structure that is a member of the core, i.e., produces results that are comparable to those found by cooperative game theory methods. We, therefore, will validate our algorithm by comparing it to a computational "brute force" mechanism for determining the core. Validation of agent-based models is generally challenging. Several methods of validation are discussed in Gore, Lynch, and Kavak (2016); Klügl (2008); Windrum, Fagiolo, and Moneta (2007); Xiang, Kennedy, Madey, and Cabaniss (2005). For our purpose, black box validation, or comparison of inputs and outputs between models, will suffice.

The brute-force algorithm, described in the next section, allows us to generate the entire core set of a hedonic game from which we can ascertain if the ABM produced a coalition structure that is a member of that core. The drawback of the brute-force method is that it can only be done for a limited number of agents due to computational requirements.

## BRUTE FORCE ALGORITHM

To be able to determine the accuracy of our heuristic algorithm, we must solve the games using another method first so that the solutions of the two approaches can be compared. To do this, we use the brute force method. A naïve or brute-force method for determining the core requires three primary steps: (1) generating every possible coalition structure; (2) generate and evaluate every possible coalition; and (3) comparing every possible coalition to each coalition structure to determine if the coalition blocks the coalition structure. The number of potential coalitions that can be formed is $2^n-1$; the number of possible coalition structures for *n = 20* agents exceeds 51 trillion, for example (Rahwan et al., 2009).

The brute-force algorithm is used for two reasons: (1) the complete core is guaranteed to be found; (2) the complete individual checks allow for simpler verification of the algorithm. The downside is that the number of agents for which the algorithm can be used is limited. The algorithm has several steps to finding the core.

1. *Generate all possible coalition structures.*

Djokić, Miyakawa, Sekiguchi, Semba, and Stojmenović (1989) presents an iterative algorithm shown in Figure 1 for generating all possible set partitions. This algorithm is executed in a Python program for coalition structure generation. The pseudocode for the algorithm is given in Figure 1.

```
setpart(n);
  initialize
    c[n]
    b[n]
    r = 1
    j = 0
    n1 = n – 1
    c[1] = 1
    b[0] = 1
    continueLoop = true

while (continueLoop)
  while ( r < n1)
    r++
    c[r] = 1
    j++
    b[j] = r
  end while
  for j = 1 to n – j
    c[n] = j
     print out c[1] to c[n]
  end for
  r = b
  c[r]++
  if c[r] > n – j then j = j - 1
  if r = 1 then continueLoop = false
end while
```

Figure 1: Pseudocode for set partition generation adapted from Djokić (Djokić et al., 1989).

2. *Generate all possible coalitions and determine the payoff for each.*

There are $2^n - 1$ possible coalition and iteratively generated using its binary representation. Agent membership, to a given coalition, is determined by a Boolean number. Each "1" bit signifies that the agent is a member of coalition *C*. The payoffs are calculated for each coalition.

In the coalition structure of a non-cooperative game, the marginal contribution is not a factor. The payoff for each agent is purely determined by which agents are a member of its coalition. For the glove game, the payoffs are either determined by the number of pairs of gloves the coalition has. The payoff division, amongst the agents, is a function of the cardinality of the agent *i*'s coalition *(C)* and the agent *i*'s utility function. The possible payoff function, for a given agent i in coalition C, is defined as follows:

$$u_i^1 = \frac{\min(\sum_{x \in C} L(x), \sum_{x \in C} R(x))}{|C|}$$

L(x) stands for the number of left gloves agent x has, and R(x) is the number of right gloves. A detailed discussion on the glove game is given below.

3. *Test each coalition structure against every coalition payoff to determine if the coalition structure is blocked.*

Determining if the coalition structure *(CS)* is blocked is a simple but computationally expensive process depicted in Figure 2. The approach requires three iterative loops. The outer loop is structured as in step 2 with the loop executing $2^n – 1$ times, with the number of each pass being converted to a binary value to represent each agent that exists in the coalition *(C)*. A second (nested) loop then iterates through every possible *CS*. If the *CS* is not blocked, a third (nested) loop iterates over the number of agents (length of the binary string) to test if the agent is a member of *C* and if the payoff of the agent in *C* is in is greater than the payoff the agent receives in *CS.* If all agents in the tested *C* can improve their position over the *CS*, then *CS* is marked as blocked.

```
CheckCorePartitions (P)
   # P is the list of partitions (coalition structure)
   # C is the list of coalitions
   # P.payoff[ ] is the list of payoffs for each player in the partition
   # C.payoff[ ] is the list of payoffs for each player in the coalition

   set P.blocked[ ] = 0
   for x = 1 to 2^N
      set n[ ] = binary(x)
      for q = 0 to length(P) – 1
         if P.blocked[q] = 0
            set P.core = TRUE
            for y = 0 to N – 1
               if n[y] = 1 then
                if C.payoff[y] <= P.payoff[y]
                   set P.core = FALSE
                   set y = N + 1
                end if
               y++
            end for
         end if
         if P.core = TRUE
            P.blocked[q] = 1
         q++
      end for
      x++
   end for
```

Figure 2: Pseudocode to determine which partitions are members of the core of a cooperative game.

4. *Store all coalition structures that are not blocked in the core file.*

A final pass is made through the coalition structures to denote that any *CS* that is not blocked is part of the core. The core of the game is static. Execution of the brute force algorithm provides a complete listing of the coalition structures that are in the core. These outputs will be used to determine the effectiveness of the ABM by providing a solution set for the ABM results to be compared.

## USE CASE – THE GLOVE GAME

Our use case is a modified version of a simple exchange economy called the glove game (Hart, 1985). This game was chosen because of its simplicity, the core is known to exist, and it can be represented as a hedonic game; however, for our purposes, we represent the glove games using players' payoff allocations as opposed to preferences for ease of understanding. In the glove game, players attempt to create pairs of gloves. Each player is endowed with initial resources – a random number of left gloves and a random number of right gloves. Players form coalitions and determine the number of pairs of gloves that are created by the group, the coalition's value. This value is then divided equally between the players to determine each player's payoff or preference for the group. Each player attempts to maximize their individual payoff.

Figure 3 is an example of the glove game with three players: *a*, *b*, and *c*. The first table shows the initial resources of each of the players. The second table enumerates each possible coalition and the utility each player in the coalition receives if that coalition is formed; the third table list all possible coalition structures and the individual values a player would receive in that coalition structure. Examining each possibility shows that player *b* receives the highest value when not combining with any other player. Without the possibility of forming a coalition with b, players *a* and *c* both obtain the greatest value by forming a coalition. Thus the core solution set is [(a,c)(b)]. This is the only coalition structure that is in the core.

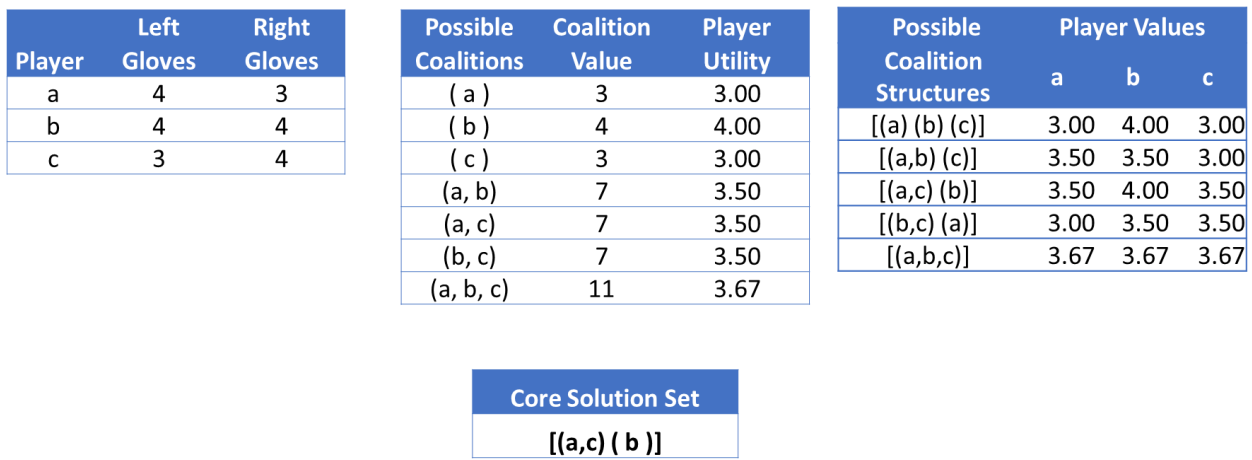

| Player | Left Gloves | Right Gloves |
|---|---|---|
| a | 4 | 3 |
| b | 4 | 4 |
| c | 3 | 4 |

| Possible Coalitions | Coalition Value | Player Utility |
|---|---|---|
| ( a ) | 3 | 3.00 |
| ( b ) | 4 | 4.00 |
| ( c ) | 3 | 3.00 |
| (a, b) | 7 | 3.50 |
| (a, c) | 7 | 3.50 |
| (b, c) | 7 | 3.50 |
| (a, b, c) | 11 | 3.67 |

| Possible Coalition Structures | Player Values | | |
|---|---|---|---|
| | a | b | c |
| [(a) (b) (c)] | 3.00 | 4.00 | 3.00 |
| [(a,b) (c)] | 3.50 | 3.50 | 3.00 |
| [(a,c) (b)] | 3.50 | 4.00 | 3.50 |
| [(b,c) (a)] | 3.00 | 3.50 | 3.50 |
| [(a,b,c)] | 3.67 | 3.67 | 3.67 |

| Core Solution Set |
|---|
| [(a,c) ( b )] |

Figure 3: Example of 3-player glove game

The glove game has had a number of applications over the years; it is also known as the shoe game (Hart & Kurz, 1983). Hart and Kurz (1983) and Hart (1985) first used the glove game to exemplify the different solution mechanisms to cooperative game theory. It has also been used in human experiments in Murnighan and Roth (1977) and Murnighan and Roth (1980). In Murnighan and Roth's first experiment, they only considered three human players and were looking at the effects of communication openness; they concluded that if a player in a weak situation controls that the flow information then they can increase the payoff they receive; this is an expected result. In their second experiment, they consider up to seven players in the "shoe" game; this time, they looked at the impacts of veto power; their experiment involved 250 participants. We believe that this paper represents the first time the glove game has been used in an agent-based model.

The ABM was designed in NetLogo, an agent-based simulation development platform (Wilensky, 1999). For game instance, each agent was randomly assigned between zero and nine left gloves and right gloves Each game involved three to nine agents; for each game size, ten unique games were designed and a complete list is given in Appendix A. Due to the stochastic nature of algorithm, each game had 50 instances run (so a total of 3500 runs were completed). Each game is executed for 100,000 timesteps to ensure convergence, if possible. In every game, each agent starts in its singleton coalition. To test whether the algorithm improved upon the original Collins and Frydenlund algorithm, their algorithm was converted to execute the glove game under the same circumstances. Both models were executed and compared to the results from the brute force algorithm to see if the agents had found a core partition.

## RESULTS AND ANALYSIS

This section discusses the results of the simulation runs. The results focus on the comparison of the effectiveness of the two algorithms (the original Collins-Frydenlund algorithm and the new one presented in this paper). Figure 4 shows the percentage of times the original Collins-Frydenlund algorithm reached a coalition structure that is a member of the core solution set (i.e., a core partition). Figure 5 shows the results of our algorithm.

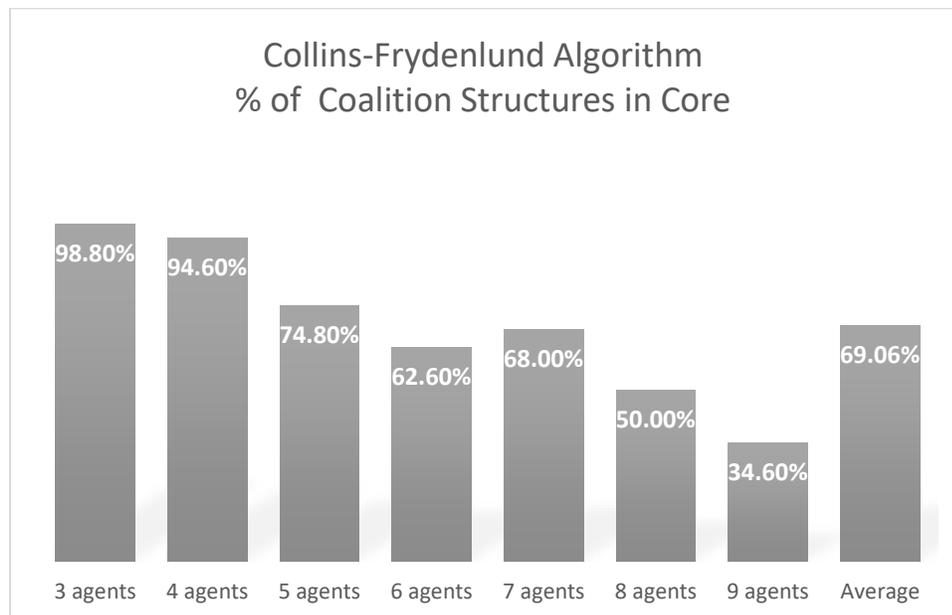

Figure 4: Percentage of times the converted Collins-Frydenlund algorithm simulation coalition structure was part of the core solution.

The new algorithm significantly outperforms the old algorithm. Also, the new algorithm has a high rate of finding a core member as its solution. Overall, 96.1% of the games resulted in finding a core member, under the new algorithm; that is, out of the 3500 games executed, the resulting coalition structure was a member of the core 3363 times. Games with three-agents, four-agents, and seven-agents always found a core member; 488 out of 500 five-agent games found a core member; 498 of the six-agent games resulted in a core coalition structure, and 499 games played with nine-agents achieved a core result.

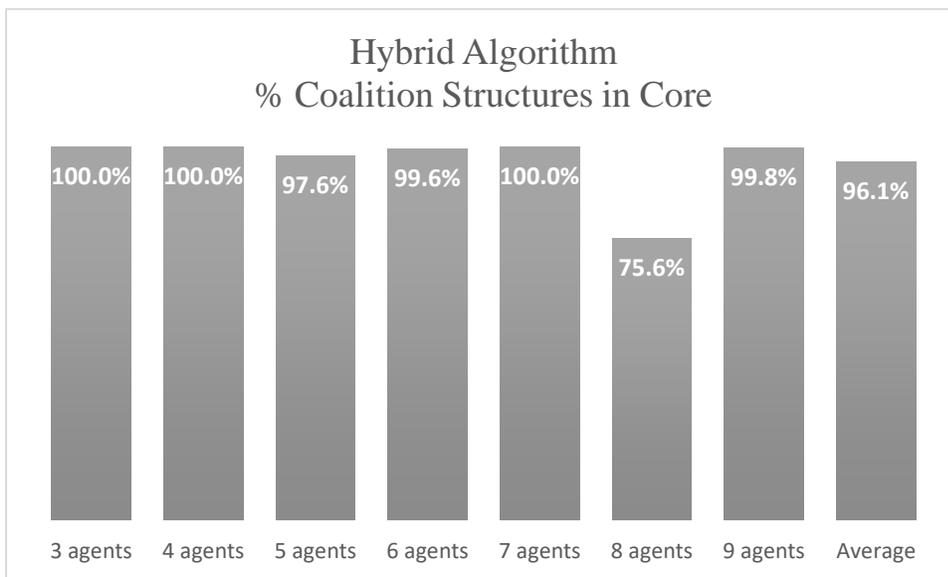

Figure 5: Percentage of times the simulation coalition structure was part of the core solution.

The only concerning result from this experiment is the eight agent games. The eight-player games, by contrast, significantly underperformed. Almost 25% of the time (122 out of 500), the eight-agent game failed to conclude with a coalition structure that could not be dominated. The details of these games are shown in Figure 6. An examination of the games that failed to reach a core solution showed that these games reached a local maximum. The algorithm used for the ABM belongs to a group known as "greedy algorithms". These algorithms are based on achieving local optimal decisions with the hope of a global solution. However, the global solution is not actually represented or considered. Therefore, it is possible for a coalition to reach a value in which it is unwilling to reconsider given the rules for a change. That is the game stalls at a local maximum. The following table, and resultant discussion, attempts to explain why this occurred with an example.

| **Players Resources** |
| --- |

| Players Resources | Player 0 | Player 1 | Player 2 | Player 3 | Player 4 | Player 5 | Player 6 | Player 7 |
|---|---|---|---|---|---|---|---|---|
| Left Gloves | 0 | 2 | 9 | 0 | 6 | 2 | 3 | 7 |
| Right Gloves | 4 | 4 | 4 | 8 | 3 | 1 | 5 | 9 |

(a)

| | Agent Coalition Payoff Values | | | | | | | |
|---|---|---|---|---|---|---|---|---|
| Core Coalitions | Player 0 | Player 1 | Player 2 | Player 3 | Player 4 | Player 5 | Player 6 | Player 7 |
| (0, 5) (1) (2, 3, 4) (6) (7) | 1 | 2 | 5 | 5 | 5 | 1 | 3 | 7 |
| (0) (1) (2, 3, 4) (5) (6) (7) | 0 | 2 | 5 | 5 | 5 | 1 | 3 | 7 |
| (0) (1,5) (2, 3, 4) (6) (7) | 0 | 2 | 5 | 5 | 5 | 1 | 3 | 7 |
| Coalition Structure Achieved | | | | | | | | |
| (0) (1,4) (2, 6) (3) (5) (7) | 0 | 3.5 | 4.5 | 0 | 3.5 | 1 | 4.5 | 7 |

(b)

Figure 6: Player coalition preference values for an eight-agent game that does not achieve a core member solution; the first agent is represented by zero which is common in computing terminology. (a) Player's resource in the game; (b) Resulting in coalition structures.

The first part (a) shows the initial resources for each of the eight agents. The core coalition section of the second part (b) is the set of coalition structures and their payoff values within the core. Obviously, each player would like to obtain the highest payoff. Part (b) also contains an example of a "sub-optimal" coalition structure that was achieved by the algorithm. To understand why a core partition was never achieved, we need to consider the linchpin coalition. The linchpin coalition in the core, in the game under consideration, consists of agents two, three, and four. A linchpin coalition is a coalition that occurs in all core partitions. Technically, the singleton coalition of agent seven is also a linchpin coalition, so is agent six's singleton group. Thus, for a core partition to evolve, under our algorithm, from a given coalition structure, the linchpin coalition must be able to form. Unfortunately, in the example given above, this is not possible from the coalition structure achieved because of the limitation of the algorithm.

The coalition consisting of agents two, three, and four represents the highest payoff value that those agents could achieve, and it would be expected that they would always be together. However, during the simulation, the coalition structure forms a coalition consisting of agents one and four prior to achieving the two, three, four combinations. As a group, neither agent one nor four improves their position by isolating themselves and agents one or four do not improve their position by joining any core member coalition structure. Therefore, agents one or four are not willing to change coalitions. Also, the coalition does not benefit from merging with any other existing coalition. Hence, under our algorithm, there is no possibility that the coalition, of player one and four, will change. In fact, there is no incentive for any coalition in the coalition structure to change, given the rules of the algorithm. This leads to a local maximum being reached.

The local maximum is not a member of the core and purely an artifact of the algorithm. This is due to the pair-wise nature of the algorithm; that is, at most, only two agents or coalitions are considered at any one time. For the linchpin coalition to form from the local maximum, the three agents, two, three and four, would need to be considered at the same time, within the algorithm step, which is not possible because they are in three different coalitions. Our heuristic algorithm is, in this respect, limited. However, the results demonstrate that, within a margin of error, it is possible to arrive at highly probable stable coalition structures which may or may not be in the core.

Why this phenomenon occurs regularly in games of eight players is not clear. We have chosen not to speculate here and leave this investigation to further research. This problem of being trapped in a local maximum could be overcome by using approaches likes simulated annealing (Van Laarhoven & Aarts, 1987), which we also leave for further research.

These results demonstrate that the ABM is effective in achieving a stable coalition structure in which the players have no incentive to move to another coalition (at least under the algorithm restrictions). However, it does not identify every member of the core set; that is, not every possible core partition was reached by the algorithm. The ABM results tended to be biassed towards one core partition over all others in the core. Figure 4 shows that only 36% of the possible core partitions were achieved. Further, there is a significant decrease in the average percentage of possible coalition structures reached when the number of players increases from 6 to 7. This is most likely due to the size of the core increasing with little to no change in the number of unique coalition structures achieved. Specifically, for the 10 games with 6 players, there are a total of 61 coalition structures in all their cores; this increases to 174 when there are 7 players. However, the number of unique coalition structures reached in the simulation is 15 and 16 for 6 and 7 players respectively.

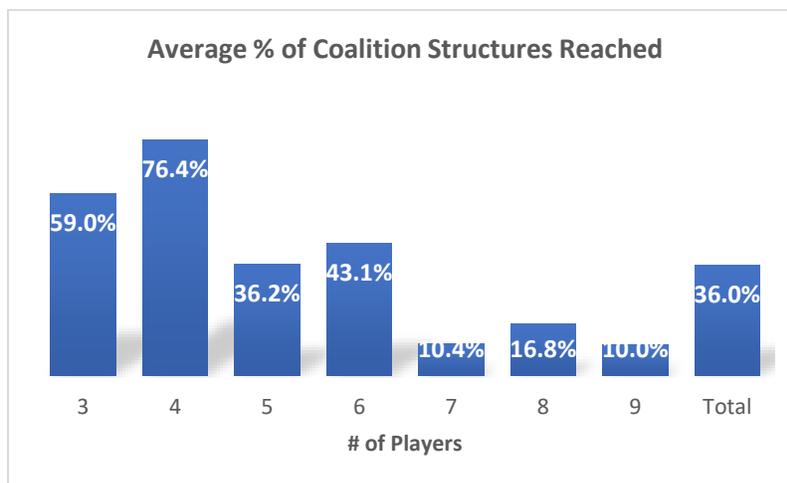

Figure 4: Percentage of core coalition structures that are reached by the simulation.

We cannot conclude that certain coalition structures cannot be reached by our algorithm because only a limited number of runs were completed for each game, i.e., fifty per game. Further research could be conducted to see if this percentage, of covered core partitions, increases significantly with an increase in the number of runs. However, if it turns out to not be the case, then another interesting direction of research is to investigate the properties of the reached core partition, for example, do they hold the trembling hand property required for a unique Nash Equilibrium to be selected in normal-form game theory (Harsanyi & Selten, 1988).The benefit of the ABM algorithm is the ability to, with computational ease, determine a coalition structure that is a core stable. The predetermined run's length of the simulation ensures that the resolution occurs within polynomial time. It is shown to have greater than a 99% effective rate overall with the lowest level exceeding 93%. The limitation of the algorithm is its narrowness in the solution set and its potential to locked into a local maximum

## CONCLUSION

There are often comparisons made between the value of game theory and simulation in social sciences (Balzer, Brendel, & Hofmann, 2001). However, agent-based simulation can be used as a complementary process to cooperative game theory. Using the game-theoretic structure and solution concepts, we designed an algorithm that produces a stable coalition structure in over 99% of the runs during our experiment, which involved the glove game with varying number of players. This was tested against a computational "brute-force" algorithm that determines the entire core, i.e., all stable coalition structure of a game.

While the brute-force method determines the complete core, it is computationally intractable for a large number of players; the ABM algorithm requires significantly less computational resources. However, there are limitations to the algorithm. The ABM solutions only accounted for 35% of the core partition solutions with a sharp decrease in this number as the number of players increased. This is most likely due to the structure of the algorithm that tends towards local maxima and only focuses on pairwise comparisons. In computer terms, our solution is a "greedy algorithm", which maximizes its value at each step thus focuses purely on exploitation over exploration of the solution space. Our algorithm mimics traditional ABM structure in that it is a bottom-up design without centralized coordination. However, it can limit the formation of a better coalition if the individuals do not immediately prosper (as demonstrated in the eight-player game example given in the Results section). One further limitation is that the algorithm will not recognize when the core is empty. Determining whether the core of a hedonic game is empty is NP-complete (Ballester, 2004). Our "greedy algorithm" will always select a value and create a coalition structure even in instances where the core is empty.

Finding core members in a hedonic game is useful in several disciplines and domains, including political science, ecology, and economics. The algorithm is intended to aid researchers that would like to apply the principles of cooperative game theory without the computation complexity. Additionally, using an ABM algorithm provides a simple means to incorporate different variables into the players' utility or preference structure without the consequence of worrying about the analytical complexity.

Future work on this algorithm will examine ways in which to reach a greater number of core members. The limited coverage of core members creates a bias and limits its use. Expanding the scope of core members achievable by the algorithm will represent a significant advancement. However, understanding why the algorithm is limited to certain types of core partitions might give insight into the nature of those core partitions. Another piece of future work is to consider more general hedonic games than the glove game presented here.

**APPENDIX A.**

| # Players (N) | Left Gloves | Right Gloves | # Players (N) | Left Gloves | Right Gloves |
|---|---|---|---|---|---|
| 3 | [3, 2, 1] | [3, 3, 2] | 7 | [1, 2, 9, 6, 1, 8, 2] | [8, 6, 6, 6, 8, 3, 0] |
| 3 | [2, 2, 2] | [1, 0, 0] | 7 | [6, 4, 8, 6, 7, 6, 5] | [4, 5, 5, 8, 8, 1, 1] |
| 3 | [4, 4, 3] | [3, 4, 4] | 7 | [1, 2, 0, 3, 5, 9, 0] | [8, 5, 6, 3, 3, 5, 8] |
| 3 | [4, 3, 1] | [3, 3, 3] | 7 | [6, 9, 9, 3, 6, 7, 4] | [5, 9, 3, 4, 9, 6, 4] |
| 3 | [3, 0, 2] | [0, 0, 3] | 7 | [2, 4, 1, 1, 7, 8, 6] | [6, 4, 0, 7, 7, 0, 0] |
| 3 | [0, 0, 3] | [3, 2, 1] | 7 | [4, 6, 2, 4, 4, 2, 0] | [0, 8, 8, 5, 8, 3, 8] |
| 3 | [0, 6, 1] | [9, 9, 0] | 7 | [3, 6, 4, 9, 7, 6, 3] | [5, 9, 3, 0, 0, 4, 8] |
| 3 | [2, 5, 8] | [4, 4, 1] | 7 | [0, 9, 8, 3, 2, 3, 3] | [6, 3, 0, 8, 2, 4, 5] |
| 3 | [7, 2, 0] | [3, 6, 4] | 7 | [8, 2, 9, 5, 4, 6, 9] | [4, 6, 3, 7, 0, 2, 3] |
| 3 | [1, 0, 4] | [0, 4, 6] | 7 | [6, 2, 6, 1, 7, 0, 8] | [5, 2, 7, 7, 3, 6, 5] |
| 4 | [0, 1, 9, 6] | [4, 5, 0, 6] | 8 | [0, 2, 9, 0, 6, 2, 3, 7] | [4, 4, 4, 8, 3, 1, 5, 9] |
| 4 | [9, 8, 8, 1] | [8, 5, 1, 0] | 8 | [7, 3, 8, 4, 0, 0, 1, 5] | [6, 7, 2, 1, 5, 2, 6, 2] |
| 4 | [3, 9, 6, 0] | [7, 3, 4, 4] | 8 | [9, 4, 6, 0, 6, 2, 4, 8] | [1, 8, 4, 0, 1, 9, 7, 8] |
| 4 | [4, 1, 6, 9] | [1, 4, 7, 2] | 8 | [9, 1, 5, 5, 5, 7, 6, 5] | [4, 6, 3, 1, 9, 5, 5, 0] |
| 4 | [2, 2, 1, 2] | [8, 0, 2, 9] | 8 | [3, 6, 1, 8, 2, 6, 6, 3] | [9, 0, 2, 7, 1, 6, 6, 7] |
| 4 | [1, 6, 2, 0] | [3, 5, 8, 8] | 8 | [0, 3, 6, 0, 1, 0, 7, 8] | [0, 0, 7, 5, 4, 1, 9, 0] |
| 4 | [4, 4, 7, 8] | [9, 0, 6, 9] | 8 | [8, 2, 3, 0, 6, 6, 2, 5] | [9, 5, 4, 1, 8, 5, 3, 5] |
| 4 | [0, 0, 7, 8] | [7, 7, 9, 5] | 8 | [7, 8, 5, 5, 4, 4, 9, 5] | [0, 8, 3, 5, 5, 6, 3, 4] |
| 4 | [4, 0, 2, 0] | [7, 8, 0, 1] | 8 | [4, 0, 8, 8, 6, 9, 5, 7] | [0, 4, 7, 7, 1, 5, 5, 9] |
| 4 | [4, 3, 1, 7] | [3, 6, 0, 1] | 8 | [8, 4, 6, 1, 1, 1, 7, 7] | [6, 3, 2, 1, 2, 6, 3, 6] |
| 5 | [7, 2, 2, 7, 0] | [6, 7, 1, 8, 6] | 9 | [4, 1, 1, 1, 6, 0, 5, 1, 4] | [0, 7, 7, 4, 2, 0, 6, 0, 0] |
| 5 | [7, 7, 1, 3, 8] | [3, 5, 1, 3, 1] | 9 | [1, 9, 8, 5, 4, 4, 7, 9, 9] | [1, 3, 9, 9, 4, 9, 9, 6, 5] |
| 5 | [7, 0, 3, 5, 3] | [1, 6, 6, 2, 7] | 9 | [5, 4, 9, 0, 7, 1, 8, 0, 8] | [4, 1, 8, 6, 6, 5, 8, 1, 7] |
| 5 | [1, 7, 1, 6, 3] | [0, 8, 0, 2, 9] | 9 | [5, 4, 6, 5, 0, 3, 4, 6, 7] | [1, 2, 2, 9, 5, 1, 3, 9, 7] |
| 5 | [4, 1, 8, 7, 2] | [6, 3, 6, 1, 1] | 9 | [1, 1, 2, 5, 2, 3, 4, 2, 8] | [9, 2, 6, 7, 4, 8, 4, 2, 0] |
| 5 | [9, 2, 5, 4, 0] | [3, 6, 6, 4, 1] | 9 | [3, 2, 5, 2, 7, 8, 1, 6, 4] | [4, 7, 6, 5, 1, 3, 4, 8, 2] |
| 5 | [2, 9, 5, 4, 5] | [8, 7, 0, 7, 4] | 9 | [2, 4, 3, 4, 5, 9, 2, 3, 2] | [7, 5, 5, 8, 1, 9, 6, 5, 3] |
| 5 | [2, 2, 2, 0, 7] | [0, 1, 8, 6, 0] | 9 | [0, 4, 5, 5, 5, 1, 7, 8, 9] | [3, 8, 2, 3, 0, 0, 3, 1, 6, 6] |
| 5 | [5, 0, 1, 0, 6] | [4, 0, 4, 7, 2] | 9 | [8, 9, 8, 4, 7, 5, 2, 8, 2] | [9, 9, 9, 5, 0, 6, 5, 2, 3] |
| 5 | [8, 0, 2, 9, 1] | [2, 8, 2, 0, 3] | 9 | [1, 4, 1, 4, 8, 0, 5, 0, 9] | [5, 0, 8, 6, 4, 2, 8, 0, 0] |
| 6 | [9, 5, 4, 6, 1, 8] | [3, 8, 9, 3, 5, 7] | | | |
| 6 | [7, 9, 1, 0, 6, 5] | [3, 2, 3, 5, 1, 8] | | | |
| 6 | [9, 6, 7, 8, 5, 8] | [0, 9, 7, 2, 0, 1] | | | |
| 6 | [8, 9, 3, 5, 4, 0] | [4, 1, 8, 0, 5, 1] | | | |
| 6 | [3, 2, 4, 2, 3, 7] | [4, 0, 8, 7, 2, 7] | | | |
| 6 | [1, 0, 0, 3, 8, 7] | [6, 8, 1, 3, 3, 8] | | | |
| 6 | [4, 1, 1, 3, 8, 6] | [4, 8, 1, 1, 5, 9] | | | |
| 6 | [9, 2, 4, 4, 4, 7] | [8, 8, 1, 1, 3, 0] | | | |
| 6 | [8, 9, 5, 2, 1, 4] | [4, 3, 1, 3, 8, 1] | | | |
| 6 | [5, 2, 5, 0, 9, 0] | [6, 3, 2, 0, 3, 7] | | | |